\documentclass{ifacconf}

\usepackage{graphicx}      
\usepackage{natbib}        
\usepackage{amssymb}
\usepackage{amsmath}
\usepackage{mathtools}
\usepackage{float}
\usepackage{xcolor}

\begin{document}
\begin{frontmatter}

\title{Control of Hierarchical Networks by Coupling to 
        an External Chaotic System} 

\author[First]{Sudhanshu Shekhar Chaurasia} 
\author[Second]{Sudeshna Sinha}

\address[First]{Indian Institute of Science Education 
        and Research Mohali, Knowledge City, 
        SAS Nagar, Sector $81$, Manauli PO $140$ $306$, 
        Punjab, India (e-mail: sudhanshushekhar@iisermohali.ac.in)}
\address[Second]{Indian Institute of Science Education 
        and Research (IISER) Mohali, Knowledge City, 
        SAS Nagar, Sector $81$, Manauli PO $140$ $306$, 
        Punjab, India (e-mail: sudeshna@iisermohali.ac.in)}

\begin{abstract}           
We explore the behaviour of chaotic oscillators in hierarchical
networks coupled to an external chaotic system whose intrinsic
dynamics is dissimilar to the other oscillators in the
network. Specifically, each oscillator couples to the mean-field of
the oscillators below it in the hierarchy, and couples diffusively to
the oscillator above it in the hierarchy. We find that coupling to
{\em one} dissimilar external system manages to suppress the chaotic
dynamics of {\em all} the oscillators in the network at sufficiently
high coupling strength. This holds true irrespective of whether the
connection to the external system is direct or indirect through
oscillators at another level in the hierarchy. Investigating the
synchronization properties show that the oscillators have the same
steady state at a particular level of hierarchy, whereas the steady
state varies across different hierarchical levels. We quantify the
efficacy of control by estimating the fraction of random initial
states that go to fixed points, a measure analogous to basin
stability. These quantitative results indicate the easy
controllability of hierarchical networks of chaotic oscillators by an
external chaotic system, thereby suggesting a potent method that may
help design control strategies.

\end{abstract}


\end{frontmatter}

\section{Introduction}

Science of complex systems is an active area of research that has
helped in understanding large interactive systems ranging from
man-made systems to natural systems. Examples of such systems include
Josephson junction \cite{hadley1988phase1,hadley1988phase2}, chemical
reactions \cite{schreiber1982strange, crowley1989experimental},
semiconductor laser \cite{varangis1997frequency,hohl1997localized},
power grids \cite{menck2012topological}, neurons
\cite{golomb2001mechanisms,li2006transient}, circadian pacemaker
\cite{daan1978two} and many other biological processes. These complex
systems consist of oscillators, which when uncoupled may be periodic,
quasi-periodic or chaotic. On coupling they can exhibit a variety of
dynamical behaviour such as synchronization
\cite{boccaletti2002synchronization,arenas2008synchronization} and
oscillation quenching or suppression of oscillation.

Suppression of oscillation, in particular, is a phenomenon of special
interest for stabilization of steady states in complex systems. Such
fixed point dynamics serve as a control mechanism, for instance in
coupled lasers where the stabilization plays an important role.  On
the other hand it is relevant in the study of pathological cases of
neuronal disorders such as Alzheimer’s and Parkinson’s disease
\cite{selkoe2000toward,
  tanzi2005synaptic,caughey2003protofibrils}. Because of such
important application, research on suppression of oscillation has been
very active over the years.

In this work, we present control of hierarchical network by coupling
to an external chaotic system. This study is an extension of our
earlier work that demonstrated the \emph{suppression of chaos through
  coupling to an external chaotic system}
\cite{chaurasia2017suppression}. In that work we investigated the
behaviour of an ensemble of chaotic oscillators, coupled diffusively
{\em only to one external chaotic system}, with the external system
coupled to the oscillator group via the mean-field of the
oscillators. Remarkably, we found that this external system manages to
successfully steer a group of chaotic oscillators onto steady states,
at sufficiently high coupling strength, when it is \emph{dissimilar}
to the group of oscillators, rather than identical. The results were
independent of number of oscillators connected to the external system.

\begin{figure}
 \centering
 \includegraphics[width=0.8\linewidth]{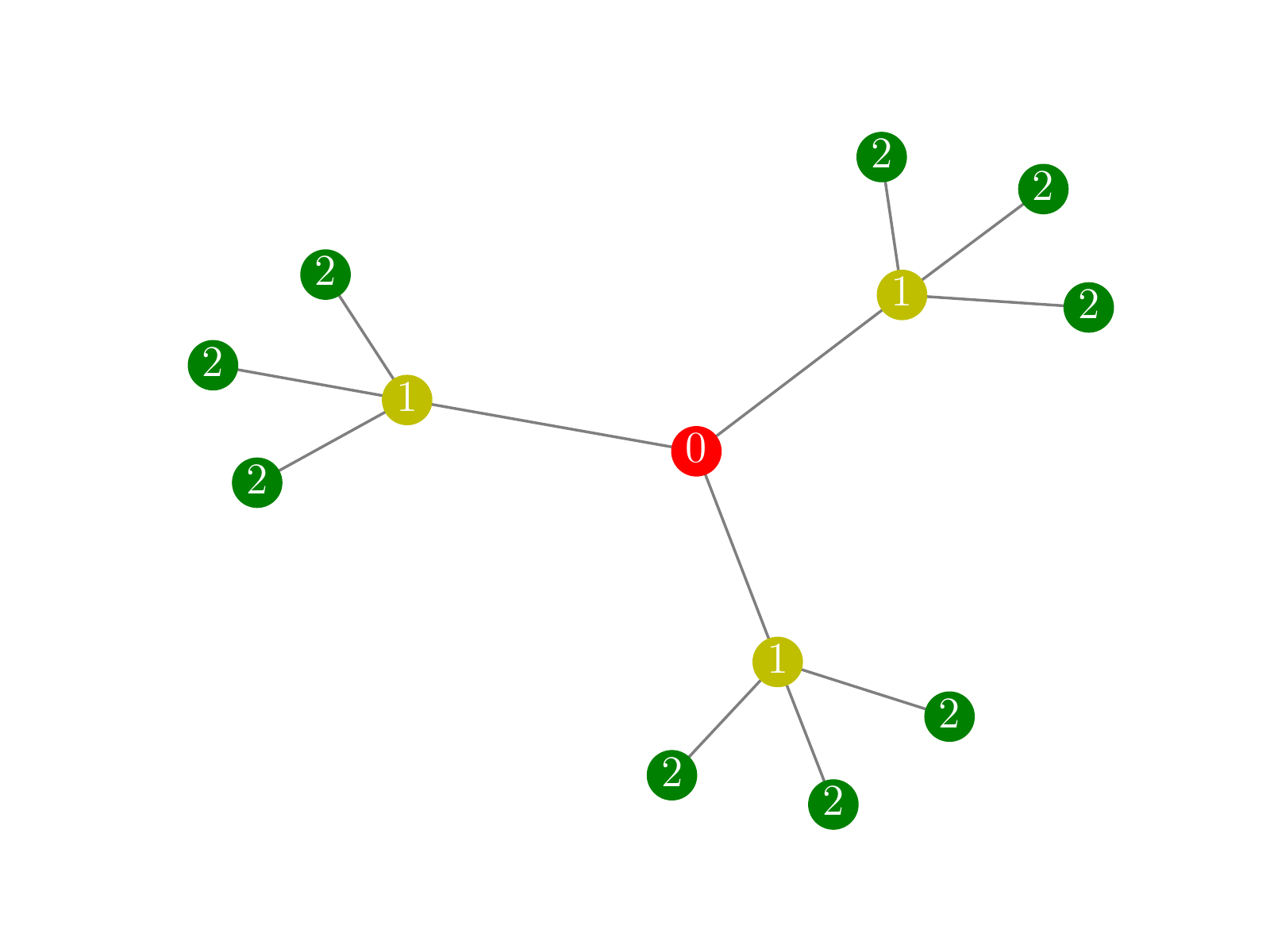}
 \caption{Schematic diagram of a hierarchical network, with two levels of hierarchy}
 \label{network:2_attachment}
\end{figure}

\begin{figure}
 \centering
 \includegraphics[width=1.0\linewidth]{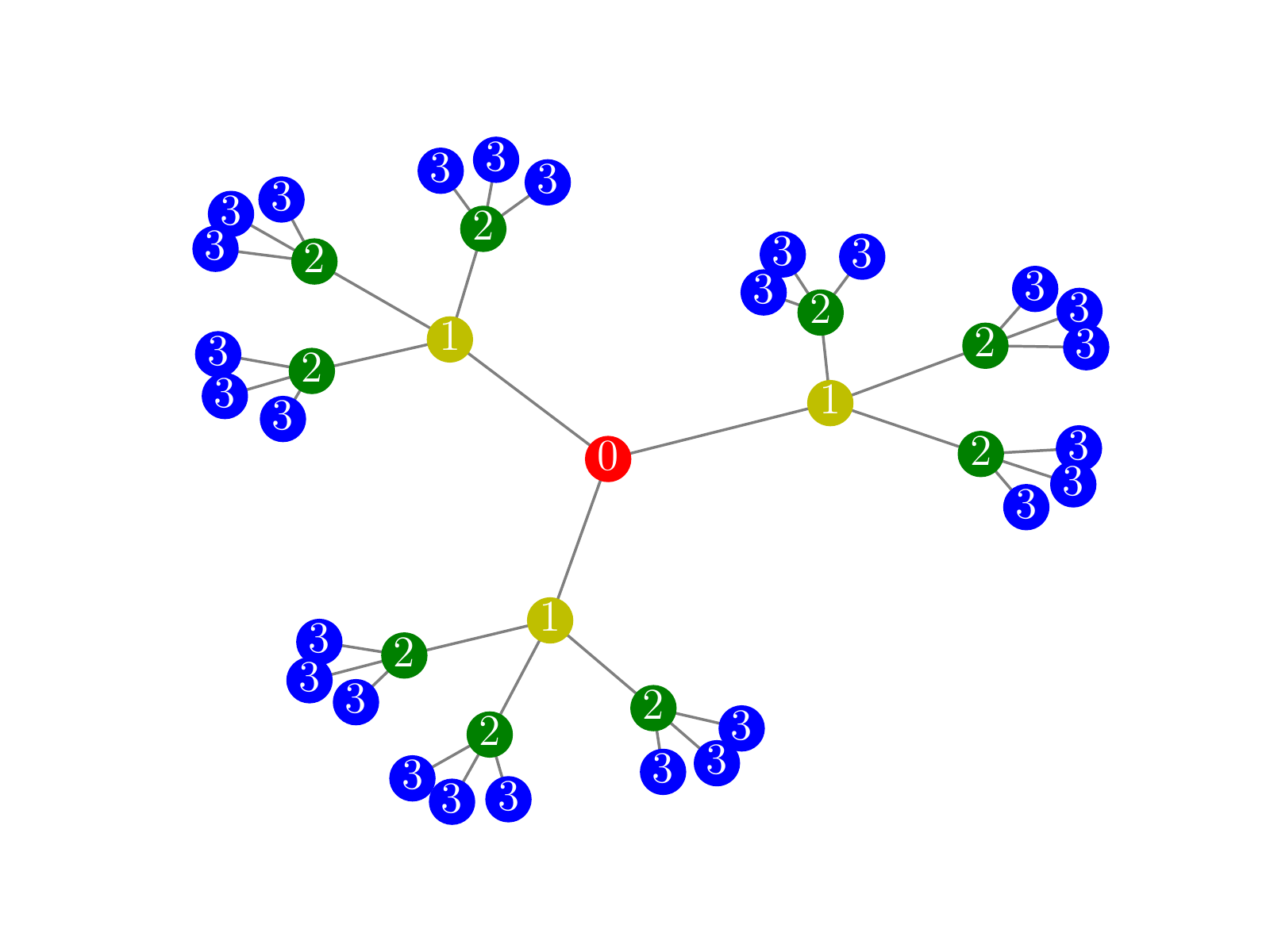}
 \caption{Schematic diagram of a hierarchical network with three levels of hierarchy}
 \label{network:3_attachment}
\end{figure}

In this work we consider a {\em hierarchical network}, with
intrinsically chaotic oscillators at different levels of the
hierarchy, and study the emergent dynamics of the network.
Figs.~\ref{network:2_attachment} and \ref{network:3_attachment} show a
schematic such networks, where level $0$ represent the external
system, which is \emph{dissimilar} to all other oscillators. The
oscillators at each level of hierarchy are connected to oscillators in
the level above and below it in the hierarchy.  For instance, for a
network with two levels (cf. Fig.~\ref{network:2_attachment}), the
oscillators at level $1$ are connected to oscillators at level $0$
(one level above in the hierarchy) and level $2$ (one level below in
the hierarchy).

Consider networks with $k$ levels of hierarchy. Examples of $k=2$ and
$k=3$ are shown schematically in Figs.~\ref{network:2_attachment} and
\ref{network:3_attachment}.  The number of oscillators at different
levels of the hierarchy is denoted by $N_l$, with level $0$ of the
network having {\em one} oscillator (i.e. $N_0 = 1$). We denote the
state variables of the $i^{\rm th}$ oscillator at level $l$ of the
hierarchical network to be: $\{x_i^{(l)}, y_i^{(l)}, z_i^{(l)} \}$,
with $i= 1, \dots N_l$. Since the $l=0$ level of the hierarchy is the
external system, we will denote it by the special symbols: $\{x_{ext},
y_{ext}, z_{ext} \}$. The mean-field of the $x$-variable of the
oscillators at different levels of the hierarchy is given by 
$\langle x^{(l)} \rangle = \frac{1}{N_l} \sum x_i^{(l)}$, 
where the sum runs over the $N_l$
oscillators at the level. Here we consider networks with $2$ and $3$
levels of hierarchy.

Now we describe the dynamics of the oscillators at different levels in 
hierarchical network. First we give the coupling form of the 
external system:

\begin{flalign}
    \label{rossler_group}
    \frac{d x_{ext}}{dt}&=f_{ext}(x_{ext},y_{ext},z_{ext}) + \varepsilon (\langle x^{(1)} \rangle - x_{ext}) & \nonumber \\
    \frac{d y_{ext}}{dt}&= g_{ext}(x_{ext},y_{ext},z_{ext}) & \\
    \frac{d z_{ext}}{dt}&= h_{ext}(x_{ext},y_{ext},z_{ext}) & \nonumber
\end{flalign}

The coupling of the oscillators at level $l=1$ of the hierarchy is given as follows:

\begin{flalign}
    \label{rossler_group}
    \frac{d x_i^{(1)}}{dt}&=f (x_i^{(1)},y_i^{(1)},z_i^{(1)}) \nonumber\\
    & \ \ \ \ \ + \varepsilon [ ( \langle x^{(2)} \rangle - x_i^{(1)} ) + (x_{ext} - x_i^{(1)}) ] & \nonumber \\
    \frac{d y_i^{(1)}}{dt}&= g (x_i^{(1)},y_i^{(1)},z_i^{(1)}) & \\
    \frac{d z_i^{(1)}}{dt}&= h (x_i^{(1)},y_i^{(1)},z_i^{(1)}) & \nonumber
\end{flalign}

The coupling of the oscillators at the intermediate levels $l=2, \dots k-1$ of the hierarchy is given as follows:

\begin{flalign}
    \label{rossler_group}
    \frac{d x_i^{(l)}}{dt}&=f (x_i^{(l)},y_i^{(l)},z_i^{(l)})  \nonumber\\
    & \ \ \ \ \ + \varepsilon [ ( \langle x^{(l+1)} \rangle - x_i^{(l)} ) + (x_j^{(l-1)} - x_i^{(l)}) ] & \nonumber \\
    \frac{d y_i^{(l)}}{dt}&= g (x_i^{(l)},y_i^{(l)},z_i^{(l)}) & \\
    \frac{d z_i^{(l)}}{dt}&= h (x_i^{(l)},y_i^{(l)},z_i^{(l)}) & \nonumber
\end{flalign}
where $j$ is the index of the node at level $l-1$ coupled to the
oscillator at level $l$. So each oscillator at level $l$ of the
hierarchy couples via the mean-field of the oscillators below it in
the hierarchy (i.e. at level $l+1$) and diffusively to the parent node
at level $l-1$.

The coupling of the oscillators at the last level $l=k$ of the
hierarchy is given as follows:

\begin{flalign}
    \label{rossler_group}
    \frac{d x_i^{(k)}}{dt}&=f (x_i^{(k)},y_i^{(k)},z_i^{(k)}) + \varepsilon ( x_j^{(k-1)} - x_i^{(k)} ) & \nonumber \\
    \frac{d y_i^{(k)}}{dt}&= g (x_i^{(k)},y_i^{(k)},z_i^{(k)}) & \\
    \frac{d z_i^{(k)}}{dt}&= h (x_i^{(k)},y_i^{(k)},z_i^{(k)}) & \nonumber
\end{flalign}
where $j$ is the index of the node at level $l=k-1$ coupled to the oscillator at level $l=k$.

The coupling strength is denoted by $\varepsilon$. 

In this work, we consider a Lorenz system in the chaotic region as the
external system, i.e. at level $0$ of the hierarchical network,
described by the dynamical equations:

\begin{flalign}
    \label{lorenz}
    f_{ext} (x,y,z) &= \sigma \ (y - x) & \nonumber \\
    g_{ext} (x,y,z) &= x \ (r - z) \ - \ y & \\
    h_{ext} (x,y,z) &= x \ y \ - \ \beta \ z & \nonumber
\end{flalign}

All other oscillators of the network are R{\"o}ssler 
oscillators in the chaotic region, with dynamical 
equations:

\begin{flalign}
    \label{rossler_group}
    f(x,y,z) &= -(\omega + \delta (x^2+y^2)) \ y - z & \nonumber \\
    g(x,y,z) &= (\omega + \delta (x^2+y^2)) \ x + a \ y& \\
    h (x,y,z) & = b + z (x-c)& \nonumber
\end{flalign}

Specifically, we consider the parameters of the Lorenz system to be
$\sigma=10.0$, $\beta=8.0/3.0$, and $r=25.0$ in Eq.\ref{lorenz} and
the parameters of the R{\"o}ssler oscillators to be $\omega=0.41$,
$\delta=0.0026$, $a=0.15$, $b=0.4$ and $c=8.4$ in
Eq.\ref{rossler_group}. These parameter sets ensure that each
oscillator is in the {\em chaotic} region when uncoupled. So in this
hierarchical network, the external system at level $l=0$ is a chaotic
Lorenz system, while the rest of the oscillators at levels $l \ne 0$
are chaotic R{\"o}ssler oscillators.


\section{Emergent Controlled dynamics}

\begin{figure}[H]
 \centering
 \includegraphics[width=1.0\linewidth]{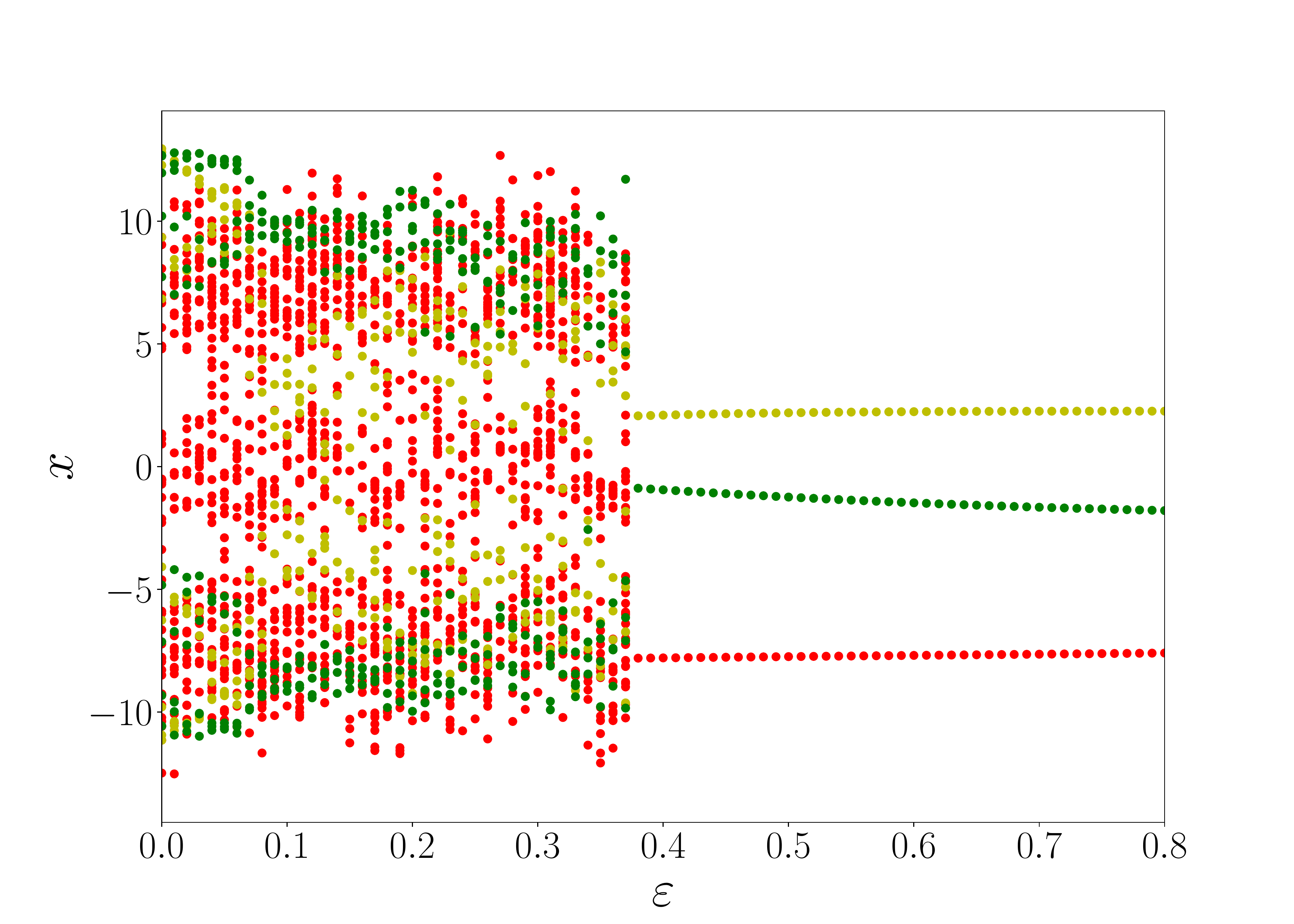}
 \caption{Bifurcation diagram, with respect to coupling strength
   $\varepsilon$, of one representative oscillator from the two levels
   of hierarchy ($l=1, 2$) and the external system ($l=0$). Colours
   correspond to the hierarchy level $l$ of oscillators, as given in
   Fig.~\ref{network:2_attachment}. In all bifurcation diagrams
   (\emph{including ones below}), we show the $x$-variable of the
   Poincare section of the phase curves of the oscillators at
   $y=y_{mid}$, where $y_{mid}$ is the mid-point along $y$-axis of the
   span of the oscillator.}
 \label{bif:second_scheme_2_attachment}
\end{figure}

First we present results for a network with two levels of hierarchy
(cf. Fig.\ref{network:2_attachment}). Fig.~\ref{bif:second_scheme_2_attachment}
shows the bifurcation diagram of the external Lorenz system and a
representative R{\"o}ssler oscillator at level 1 and 2 of the hierarchical
network. We find that the oscillators at {\em all levels of hierarchy}
can be controlled to a fixed point, at sufficiently high coupling
strength, by an external chaotic Lorenz system. The results are
unchanged on increasing the number of oscillators at level 2
(i.e. increasing $N_2$), indicating that the control is independent of
system size.

\begin{figure}[H]
 \centering
 \includegraphics[width=1.0\linewidth]{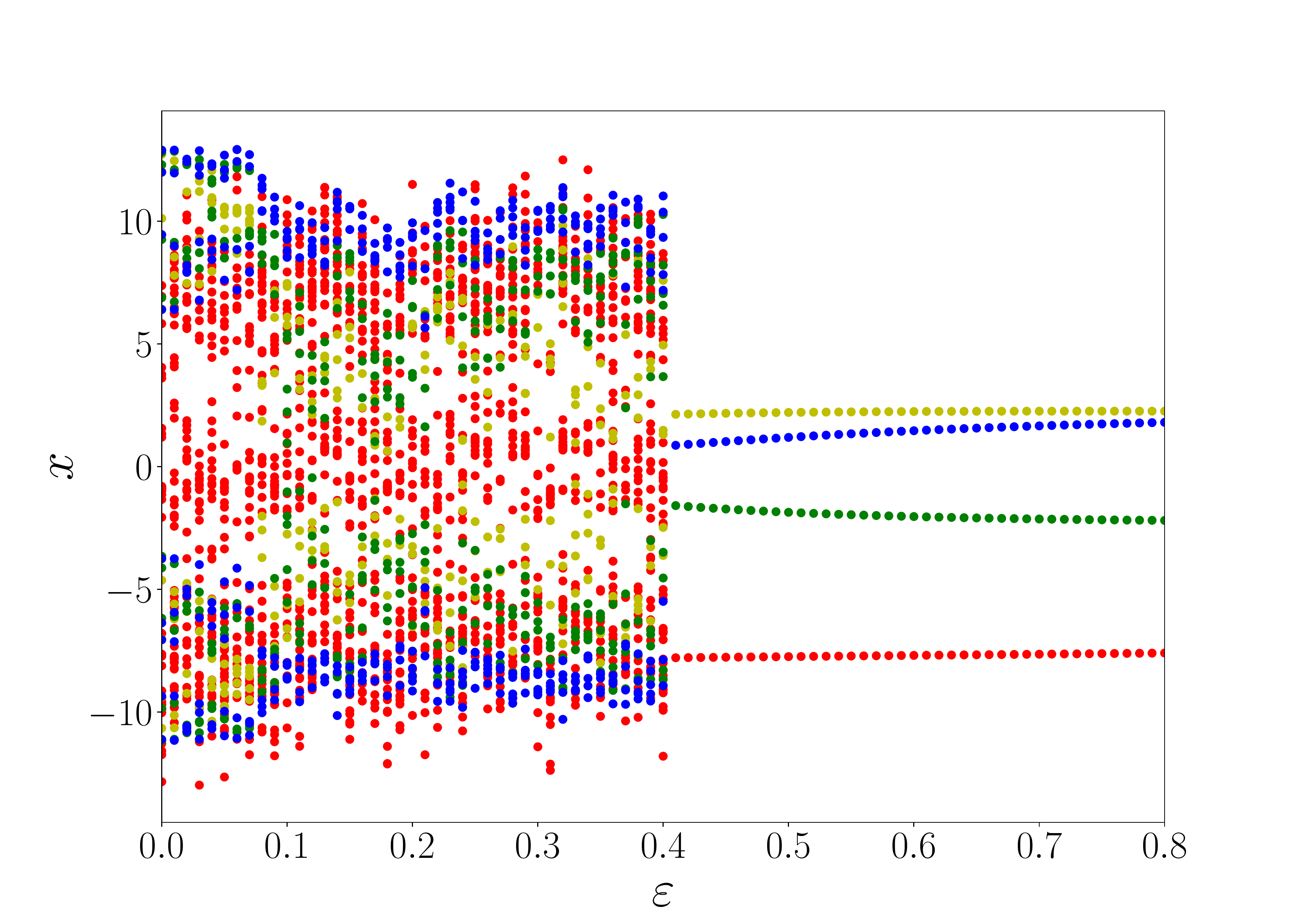}
 \caption{Bifurcation diagram, with respect to coupling strength
   $\varepsilon$, of one representative oscillator from the
   three levels of hierarchy, and the external system. Colours correspond
   to the hierarchy level of the oscillators, as given in
   Fig.~\ref{network:3_attachment}. }
 \label{bif:second_scheme_3_attachment}
\end{figure}

Now we check the generality of our results, by investigating a network
with an additional level of hierarchy, namely three levels of
hierarchy
(cf. fig\ref{network:3_attachment}). Fig.~\ref{bif:second_scheme_3_attachment}
shows the bifurcation diagram of one of the representative oscillators
from each level of hierarchy. The emergent behaviour is same for any
number of oscillators attached to level 2, forming level 3 of
hierarchy. At very low coupling strength, all the oscillators yield
their intrinsic chaotic dynamics. Increasing the coupling strength
distorts the dynamics, with distortion being different at different
levels of hierarchy. When coupling strength is sufficiently high
($\varepsilon \sim 0.4$), there is sudden transition from oscillations
to fixed points. These fixed points are different at each level of
hierarchy. However note that all the oscillations are suppressed at
exactly the same steady state for all oscillators at a particular
level in the hierarchy.

\begin{figure}[H]
 \centering
 \includegraphics[width=1.0\linewidth]{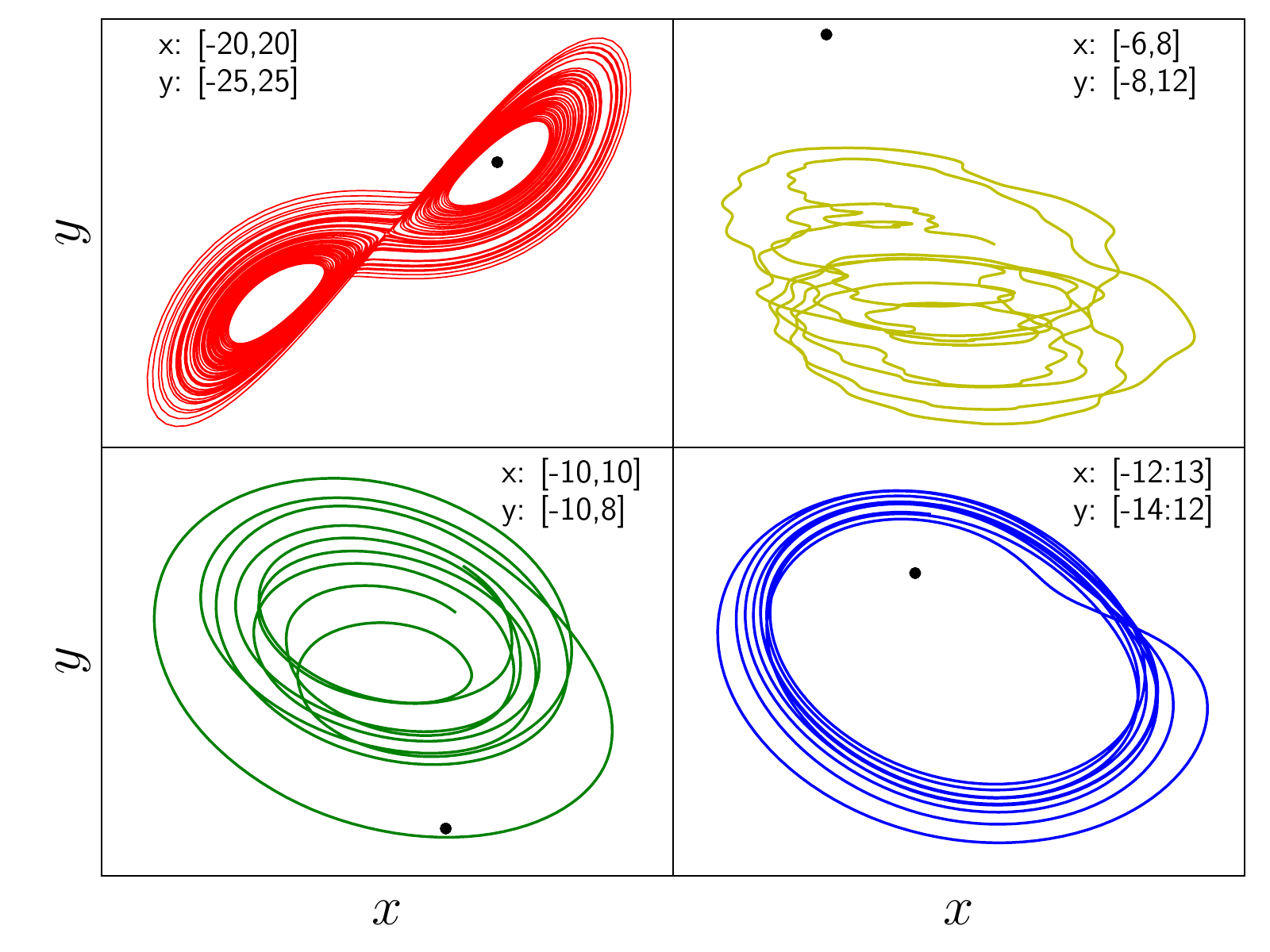}
 \caption{Phase portraits of the external system and one
   representative oscillator from each level of hierarchy. The
   coloured phase portraits are obtained for $\varepsilon=0.2$, and
   the black dots are the fixed points obtained for $\varepsilon=0.6$.
   Different colours of the phase portraits correspond to the
   hierarchy level of oscillators, as given in
   Fig.~\ref{network:3_attachment}. }
 \label{phase_diag:second_scheme_3_attachment}
\end{figure}

Fig.~\ref{phase_diag:second_scheme_3_attachment} shows the projection
of phase portraits on $x-y$ plane of one representative oscillator
from each level of hierarchy at intermediate coupling strength
$\varepsilon=0.2$. The colours correspond to the hierarchy level as
illustrated in the schematic diagrams. It is evident that the
oscillators at the level nearest to the external system are most
distorted. As the oscillators move away from the external system
($l=0$) in the level of hierarchy, the oscillations become close to
their intrinsic dynamics. The black dots in the figure show the fixed
points obtained at high coupling strength ($\varepsilon=0.6$).

\section{Synchronization}

Now we study the synchronization properties of the system in the case
of a network with three hierarchical levels. It is evident from
Fig.~\ref{bif:second_scheme_3_attachment} that the oscillators at each
level of hierarchy attain different fixed points, i.e. there is no
synchronization across different levels, though there is
synchronization within a level. here we calculate the advent of this
synchronization within a level of hierarchy in the network, as a
function of coupling strength.

\begin{figure}[H]
 \centering
 \includegraphics[width=1.0\linewidth]{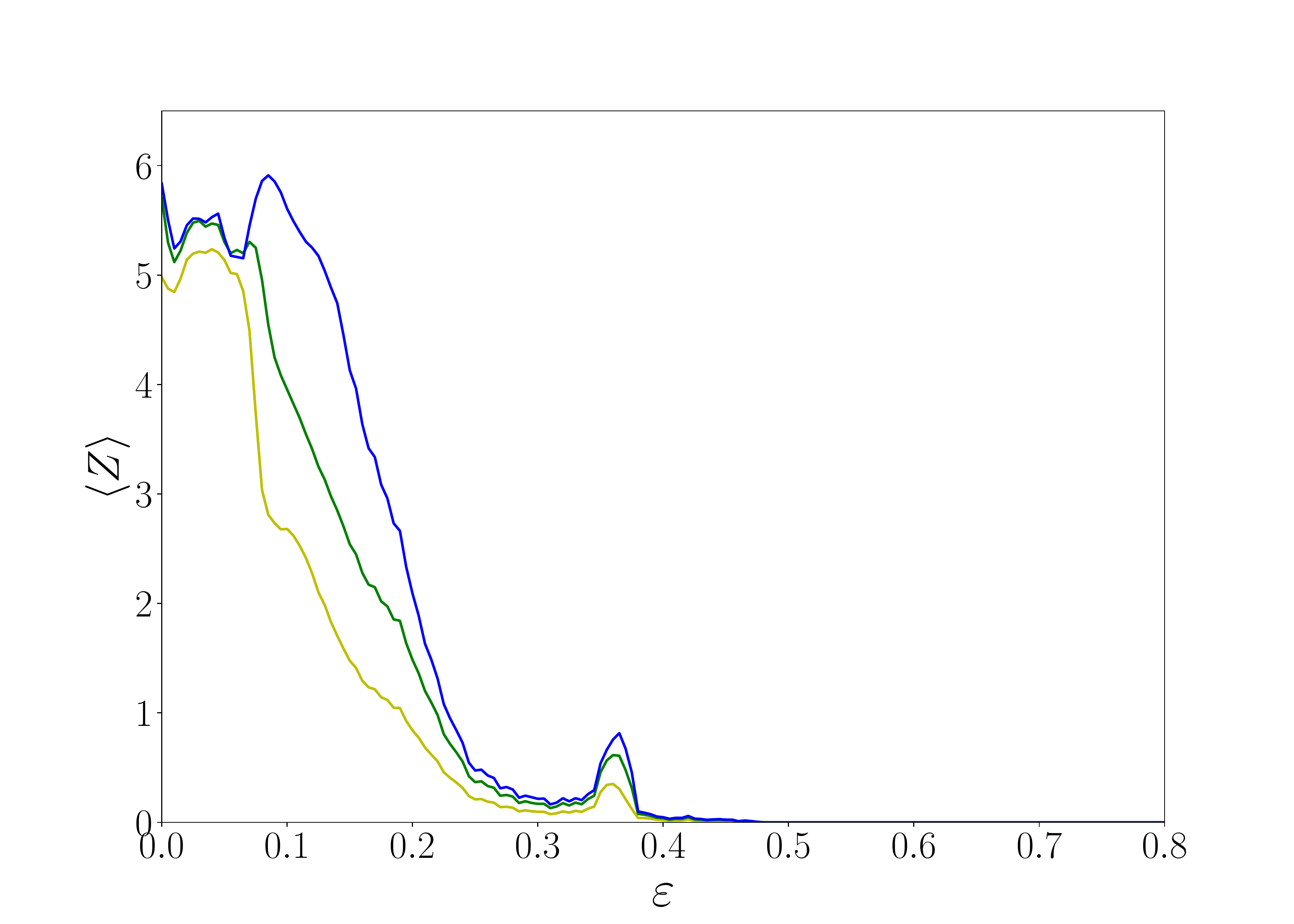}
 \caption{Synchronization error $\langle Z\rangle$ of the R{\"o}ssler 
 oscillators, with respect to coupling strength, averaged over 
 1000 initial conditions. Different colours correspond to the 
  hierarchy level of oscillators as shown in 
  Fig.~\ref{network:3_attachment}. }
 \label{sync:second_scheme_3_attachment}
\end{figure}

We calculate the synchronization error of the oscillators at 
particular levels, averaged over time $T$, given by:
\begin{equation}
 Z=\frac{1}{T} \sum_t \sqrt{(\bar{x^2})_t-({\bar{x}}^2)_t}
\end{equation}

where $(\bar{x})_t=\frac{1}{N_l}\sum_{i=1}^{N_l} x_i^{(l)}$ and
$(\bar{x^2})_t=\frac{1}{N_l}\sum_{i=1}^{N_l} (x^{(l)}_i)^2$ are the
average value of $x$ and $x^2$ of the oscillators, at an instant of
time $t$, at level $l$ of the hierarchy. Further we average $Z$ over
different initial states to obtain an ensemble averaged
synchronization error $\langle Z \rangle$.

Fig.~\ref{sync:second_scheme_3_attachment} shows the synchronization
error ($\langle Z \rangle$) with respect to coupling strength. It is
evident that oscillators within each level of hierarchy are
synchronized when the oscillations are suppressed, i.e oscillators at
the same level evolve to the same fixed points, even though there is
{\em no direct coupling between them}.

\section{Basin Stability of the Spatiotemporal Fixed Points}

Suppression of oscillations is achieved by each oscillator in the
network at the same coupling strength, i.e. control to fixed point
occurs at the same critical coupling strength $\varepsilon_c$ for all
oscillators at all levels of hierarchy. However, the value
$\varepsilon_c$ may vary with initial conditions.

We now quantify the efficacy of control to steady states by uniformly
sampling a large set of random initial conditions and estimating the
fraction $\text{BS}_{fixed}$ of initial states attracted to
spatiotemporal fixed points. This measure is analogous to recently
used measures of basin stability \cite{menck2013basin} and indicates
the size of the basin of attraction for a spatiotemporal fixed point
state. $\text{BS}_{fixed}\sim 1$ suggests that the fixed point state
is {\em globally attracting}, while $\text{BS}_{fixed}\sim 0$
indicates that almost no initial states evolve to stable fixed states.

\begin{figure}[H]
 \centering
 \includegraphics[width=1.0\linewidth]{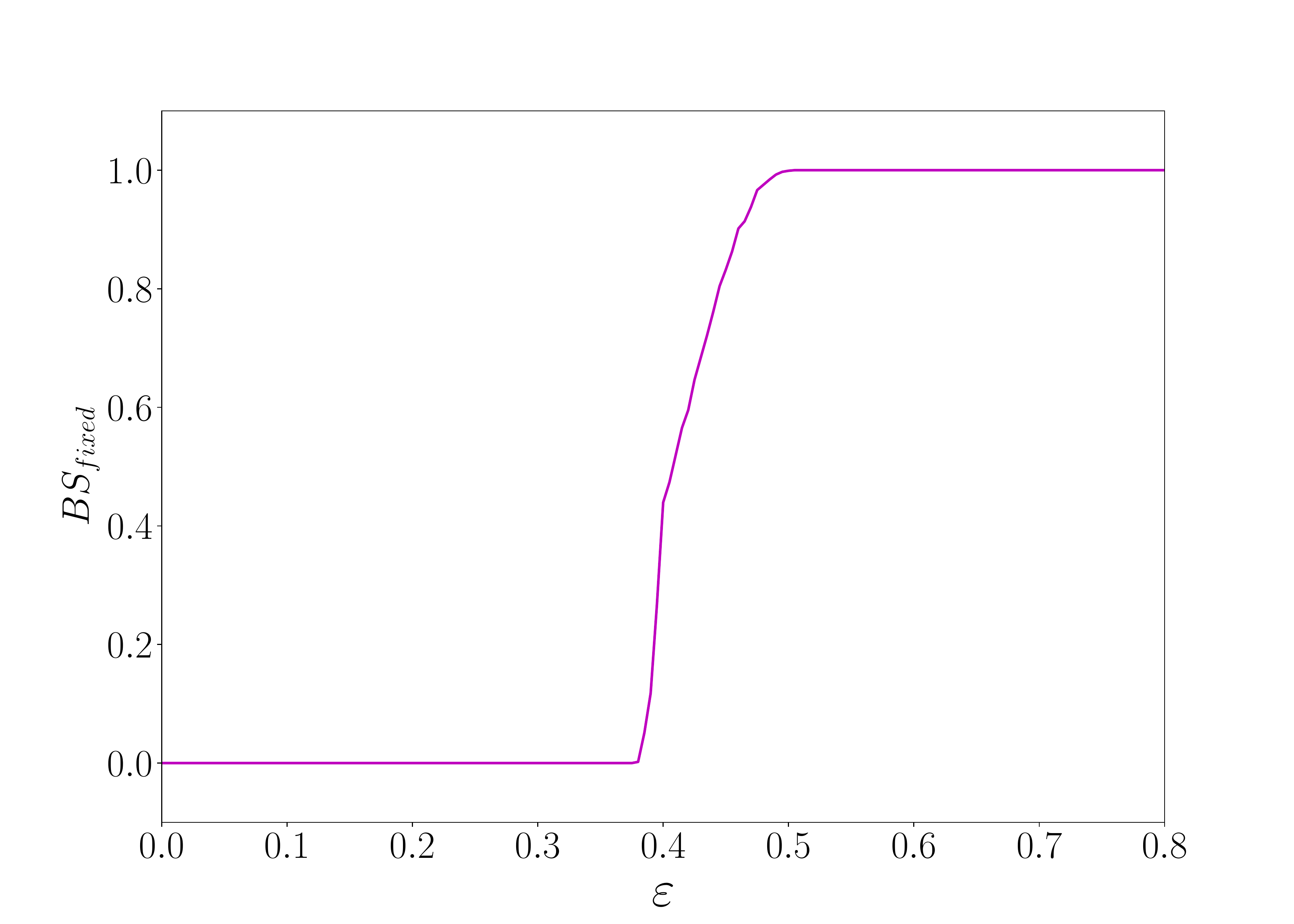}
 \caption{Dependence of the fraction of initial conditions attracted
   to the spatiotemporal fixed point, $\text{BS}_{fixed}$, on the
   coupling strength $\varepsilon$, averaged over 5000 initial
   conditions.}
 \label{basin:second_scheme_3_attachment}
\end{figure}

We show the variation of $\text{BS}_{fixed}$ as a function of coupling
strength (cf. Fig.~\ref{basin:second_scheme_3_attachment}).  For
$\varepsilon_c<0.38$, system does not evolve to the fixed point state
for any initial condition, while for $\varepsilon_c>0.5$ the network
gets attracted to a spatiotemporal fixed point for all initial
conditions. In the intermediate range of coupling strengths, the
network has a finite probability to get attracted to the
spatiotemporal fixed point state from a generic random initial state,
as indicated by $\text{BS}_{fixed} > 0$.

\section{Conclusion}

We investigated the behaviour of an ensemble of chaotic oscillators
coupled in a hierarchical network, where at the zeroth level of the
hierarchy we have {\em one chaotic external system that is dissimilar
  to the rest of the oscillators in the network}.  We have shown that
at sufficiently high coupling strengths, this one external system can
{\em suppress the oscillations at all levels of the
  hierarchy}. Therefore we have demonstrated a potent method to
control chaotic oscillators in a hierarchical network to steady
states, using one external dissimilar chaotic system.

\bibliography{ifacconf}
\end{document}